\documentclass[preprint,showpacs,preprintnumbers,amsmath,amssymb,prl]{revtex4}
\usepackage{graphicx}
\usepackage{latexsym}
\begin{document}
\date{\today}
\title{Detrending Moving Average variance: \\ a derivation of the scaling law}

\author{Sergio Arianos}
\email{arianos@infn.to.it} \altaffiliation[Permanent address:
]{Theoretical Physics Department, Universit\'a di Torino, Italy}
\author{Anna Carbone}
\email{anna.carbone@polito.it}
\homepage{http://www.polito.it/noiselab} \affiliation{
 Dipartimento di Fisica, Politecnico di Torino, Corso Duca degli
Abruzzi 24, I-10129 Torino, Italy \\ }

%

\begin{abstract}

The Hurst exponent  $H$ of long range correlated series can be
estimated by means of the Detrending Moving Average (DMA) method.
A computational tool defined within the algorithm is the
generalized variance $ \sigma_{DMA}^2={1}/{(N-n)}\sum_i
[y(i)-\widetilde{y}_n(i)]^2\:$, with $\widetilde{y}_n(i)=
{1}/{n}\sum_{k}y(i-k)$ the moving average, $n$ the moving average
window and $N$ the dimension of the stochastic series $y(i)$. This
ability relies on the property of $\sigma_{DMA}^2$ to scale as
$n^{2H}$. Here, we analytically show that $\sigma_{DMA}^2$ is
equivalent to $C_H n^{2H}$ for $n\gg 1$ and provide an explicit
expression for $C_H$.
\end{abstract}

\maketitle
\section{Introduction} Long-memory stochastic processes are ubiquitous in fields
as different as condensed matter, biophysics, social science,
climate change, finance \cite{Mantegna,Willinger,Lam, Ferreira}.
The development of methods able to quantify the statistical
properties and, in particular, to extract the Hurst exponent of
long-range correlated signals continue therefore to draw the
attention not only of the physicist community
\cite{Feder,Hurst,Mandelbrot,Peng,Vandewalle,Ivanova,Rangarajan,Heneghan,Alessio,Carbone,Dimatteo}.
For long-memory correlated processes, the value of the Hurst
exponent $H$ ranges from $0<H <0.5$  and from $0.5<H<1$  for
negative and positive persistence respectively; $H = 0.5$ is found
in fully uncorrelated signals. Several techniques have been
proposed in the literature to study the scaling properties of time
series. We limit ourselves to mention here only a few of them such
as the seminal work by Hurst on rescaled range statistical
analysis (R/S), the modified R/S analysis, the multi-affine
analysis, the detrended fluctuation analysis (DFA), the
periodogram regression (GPH) method, the (m,k)-Zipf method, the
detrended moving average analysis (DMA). The challenge is to get
the Hurst exponent $H$, that is related to the fractal dimension
$D=2-H$, by means of more and more accurate and fast algorithms.
The methods of extraction of the scaling exponents from a random
signal exploit suitable statistical functions of the series
itself. The rescaled range statistical analysis (R/S) was first
introduced by Hurst to describe the long-term dependence of water
levels in rivers and reservoirs. It provides a sensitive method
for revealing long-run correlations in random processes. This
analysis can distinguish random time series from correlated time
series and gives a measure of a signal ''roughness''.    The
 Detrended Fluctuation Analysis (DFA) is a very popular scaling
technique to estimate the power-law correlation exponents of
random signals in the time domain. Recently, a method called
Detrended Moving Average (DMA) technique for the analysis of the
persistence has been proposed \cite{Alessio}. The striking
difference between the DMA and the previously proposed (R/S, DFA)
methods
 is that  a division of the series
in boxes is not needed by the DMA algorithm. The scaling property
is obtained by using  a simple continuous function: the moving
average. This fact makes the DMA algorithm highly efficient from
the computational point of view. Many properties of the DMA have
been studied and applications have been demonstrated
\cite{Carbone}. \par In this work, we provide an analytical
derivation that the DMA variance scales as a power law with
exponent ${2H}$, i.e. $\sigma_{DMA}^2 \sim C_H n^{2H} $. The
analytical findings are compared with the simulation data obtained
by calculating the DMA variance of surrogate Fractional Brownian
motions with different values of $H$, generated by the Random
Midpoint Displacement (RMD) algorithm \cite{Voss}.

\section{METHOD}
 First we describe the main steps of the DMA algorithm. The
technique is based on the function:

\begin{subequations}
\label{DMA}
\begin{equation}
 \sigma^2_{DMA}=\frac{1}{N-n}\sum_{i=n
}^{N}\Big[y(i)-\tilde{y}_{n}(i)\Big]^2 \label{DMA:1}
\end{equation}
\begin{equation}
 \tilde{y}_{n}(i)=
\frac{1}{n}\sum_{k=0}^{n  }y(i-k) \label{MA:1}
\end{equation}

\noindent The Eq.~(\ref{DMA:1}) defines a generalized variance of
the random path $y(i)$ with respect to the moving average
$\widetilde{y}_{n}(i)$ (Eq.~(\ref{MA:1})). The function
$\widetilde{y}_{n}(i)$ is calculated by averaging the $n$-past
value in each sliding window of length $n$.  In so doing, the
reference point of the averaging process is the last point of the
window. The dynamic averaging process and the DMA algorithm can be
however referred to any point within the window, by generalizing
the  Eqs.~(\ref{DMA:1},\ref{MA:1}) as follows:

\begin{equation}
 \sigma^2_{DMA}=\frac{1}{N- n}\sum_{i=n (1-\theta) }^{N-
 n\theta}\Big[y(i)-\tilde{y}_{n}(i)\Big]^2 \label{DMA:2}
\end{equation}
\begin{equation}
 \tilde{y}_{n}(i)=
\frac{1}{n}\sum_{k=- n\theta}^{n (1-\theta) }y(i-k) \label{MA:2}
\end{equation}
\end{subequations}

\noindent Upon variation of  the parameter $\theta$   in the range
$\left[ 1, 0 \right]$, the reference point of
$\widetilde{y}_{n}(i)$ is accordingly set within the moving window
$n$. In particular, consider the following three relevant cases:
(i)~$\theta=0$ corresponds to calculate $\widetilde{y}_{n}(i)$
over all the past points within the window $n$;
(ii)~$\theta=\frac{1}{2}$ corresponds to calculate
$\widetilde{y}_{n}(i)$ over $n/2$ past and $n/2$ future points
within the window $n$;
 (iii)~$\theta=1$ corresponds to calculate
$\widetilde{y}_{n}(i)$ over all the future points within the
window $n$ respectively. \par In order to calculate the Hurst
exponent of the series, the DMA algorithm is implemented as
follows. The moving average $\widetilde{y}_{n}(i)$ is calculated
for different values of the window $n$, with $n$ ranging from 2 to
a maximum value $n_{max}$ determined by the size of the series.
The $\sigma_{DMA}$, defined by the Eqs.(\ref{DMA:1}, \ref{DMA:2}),
is then calculated for all the windows $n$ over the interval
$\left[ n, N \right]$. For each $\widetilde{y}_{n}(i)$, the value
of $\sigma_{DMA}$ corresponding to each $\widetilde{y}_n(i)$ is
plotted as a function of $n$ on log-log axes. The most remarkable
property of the log-log plot is to exhibit a power-law dependence
on $n$, i.e. $ \sigma_{DMA}^2 \sim C_H n^{2H}$, allowing thus to
calculate the scaling exponent $H$ of the signal $y(i)$.

\section{DERIVATION OF THE SCALING RELATIONSHIP}

The variance defined by the Eqs.(\ref{DMA}) will be analytically
deduced in the limit $N\gg n\gg 1$ for fractional Brownian path.
The calculation will be performed for the DMA variance with moving
average $\widetilde{y}_{n}(i)$ referred to an arbitrary point
within the moving window. We will prove that:
\begin{equation}
 \sigma^2_{DMA}\sim
  C_{H}n^{2 H}\;,\qquad n\gg 1
\end{equation}

\noindent with
\begin{equation}
\label{CHdma}
  C_{H}
  =\frac{1}{2H+1}\left[\big((1-\theta)^{2H+1}+\theta^{2H+1}\big)-\frac{1}{2(H+1)(2H+1)}\right]
\end{equation}
\vskip 0.5cm \noindent By simple transformation of the Eq.
(\ref{DMA:2}), one obtains:
\[ (N-n)\sigma^2_{DMA}=\sum_{i=n-\theta n}^{N-\theta
n}y^2(i)-\frac{2}{n}\sum_{i=n-\theta n}^{N-\theta
n}y(i)\sum_{k=-\theta n}^{n-\theta n}y(i-k)\]
\begin{equation}\label{dma00} +\frac{1}{n^2}\sum_{i=n-\theta n}^{N-\theta
n}\left(\sum_{k=-\theta n}^{n-\theta n}y(i-k)\right)^2
\end{equation}
\vskip 0.5cm \noindent Let us consider each term separately. The
first term on the right hand side of the equation (\ref{dma00})
writes:
\begin{equation} \sum_{i=n-\theta n}^{N-\theta
n}y^2(i)=\sum_{i=n-\theta n}^{N-\theta
n}i^{2H}\simeq\frac{1}{2H+1}\left[(N-\theta
 n)^{2H+1}-(n-\theta n)^{2H+1}\right]
\end{equation}
\vskip 0.5cm \noindent The second term on the right hand side:
\[
-\frac{2}{n}\sum_{i=n-\theta n}^{N-\theta n}y(i)\sum_{k=-\theta
n}^{n-\theta n}y(i-k)=-\frac{2}{n}\sum_{i=n-\theta n}^{N-\theta
n}y(i)
\sum_{j=i-n+\theta n}^{i+\theta n}y(j) \\
\]
\[\simeq  -\frac{1}{2H+1}\left[(N-\theta n)^{2H+1}-(n-\theta n)^{2H+1}\right] \]
\[-\frac{1}{2(H+1)(2H+1)}\frac{1}{n}\left[N^{2H+2}-n^{2H+2}-(N-n)^{2H+2}\right] \]
\begin{equation}
+\frac{n^{2H}}{2H+1}\left[(1-\theta)^{2H+1}+\theta^{2H+1}\right](N-n)
\end{equation}
\vskip 0.5cm \noindent Finally, the third term:
\[
 \frac{1}{n^2}\sum_{i=n-\theta n}^{N-\theta n}\big(\sum_{k=-\theta n}^{n-\theta n}y(i-k)\big)^2=\frac{1}{n^2}\sum_{i=n-\theta n}^{N-\theta n}
 \big[\sum_{j=i+\theta n-n}^{i+\theta n}y(j)\big]^2
\]
\[\simeq \frac{1}{2(H+1)(2H+1)}\frac{1}{n}\big[N^{2H+2}-n^{2H+2}-(N-n)^{2H+2}\big] \]
\begin{equation}
-\frac{n^{2H}}{2(H+1)(2H+1)}(N-n)
\end{equation}
\vskip 0.5cm \noindent Summing the contributions from each term we
obtain
\begin{equation}
\label{CH}
\sigma^2_{DMA}\simeq\left[\frac{1}{2H+1}\big((1-\theta)^{2H+1}+\theta^{2H+1}\big)-\frac{1}{2(H+1)(2H+1)}\right]n^{2H}
\end{equation}
\vskip 0.5cm \noindent One can easily check that $\sigma^2_{DMA}$
takes the following expressions   for $\theta=0$ and $\theta=1$
and for $\theta=1/2$:
\begin{itemize}

\item $\theta=0~$ or  $\theta=1~$; i.e. when the moving average is referred to the last or to the first point of the
window, it is
\begin{equation}
\label{CH0}
 C_H\simeq\frac{1}{2(H+1)} \end{equation}
\item $\theta=\frac{1}{2}$~: the average is referred to the center of the window
\begin{equation}
\label{CH05}
C_H\simeq\big[\frac{1}{2(H+1)}-\frac{1-2^{-2H}}{2H+1}\big]
\end{equation}

\end{itemize}
\vskip 0.5cm The above calculations have been performed for a
fractional brownian motion with variance $\sigma^2=t^{2H}$. It can
be easily shown that for the general case of a fractional brownian
motion with variance $\sigma^2=D_H t^{2H}$,
 the Eq.~(\ref{CH}) will asymptotically behave
as $ \sigma_{DMA}^2=D_H C_H n^{2H}\;. $
\section{RESULTS AND DISCUSSION} In this section, the values of $C_H$ obtained by calculating the DMA variance of
artificial fractional Brownian motions are compared with those
calculated using the Eq.~(\ref{CH}).

\par
\begin{figure}
\includegraphics[width=9.5cm,height=8cm,angle=0]{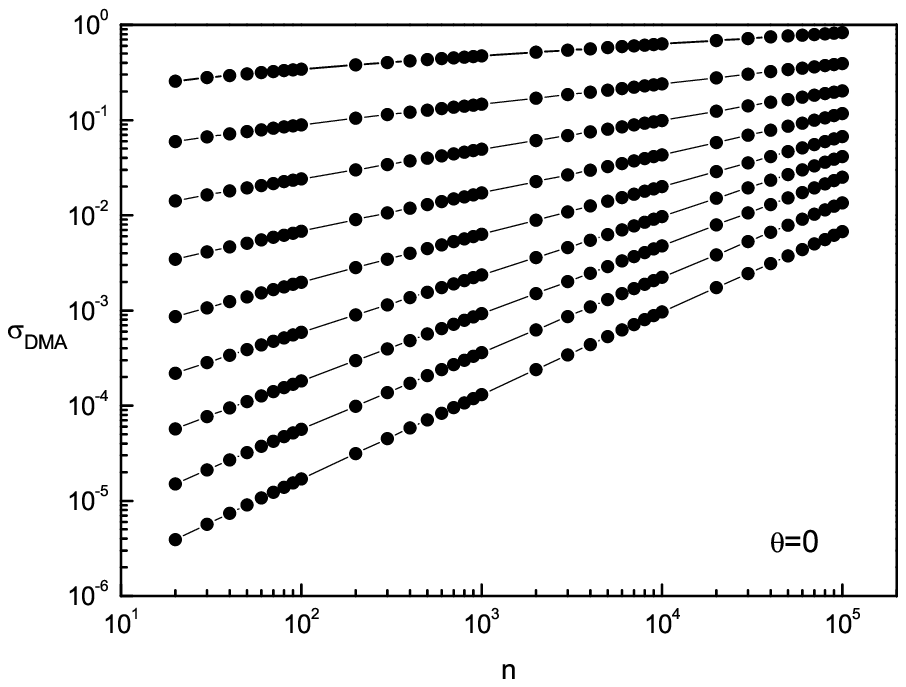}
\includegraphics[width=9.5cm,height=8cm,angle=0]{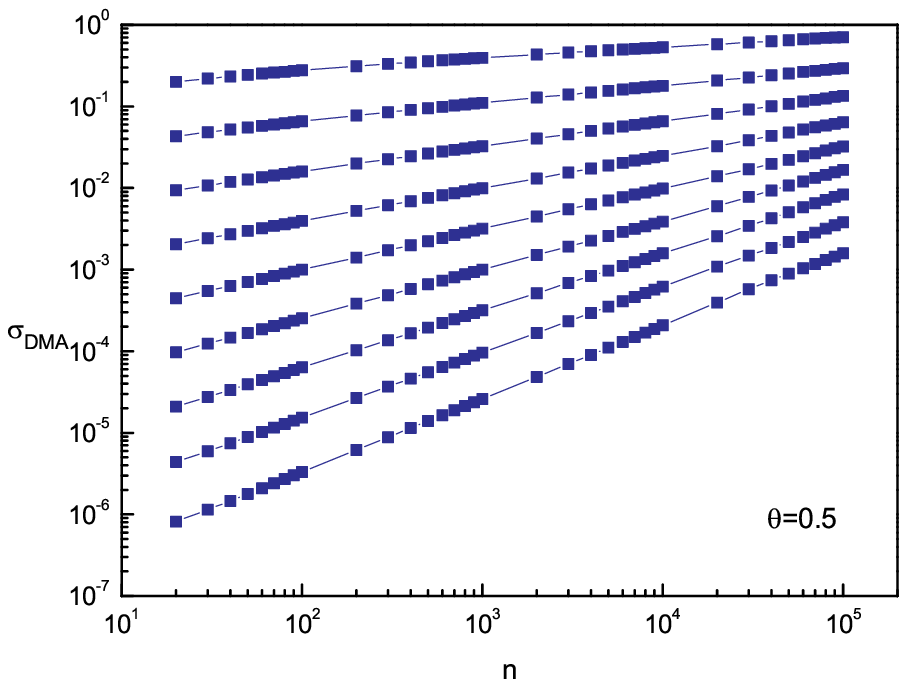}
\includegraphics[width=9.5cm,height=8cm,angle=0]{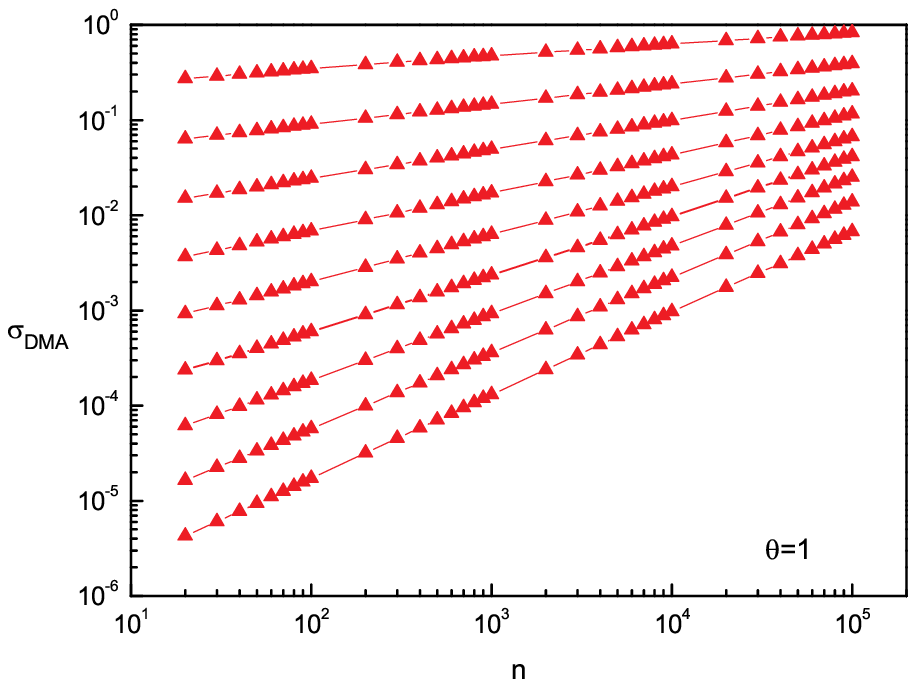}
\caption{\label{dma0} Log-log plot of the function $\sigma_{DMA}$
defined by the Eq. \ref{DMA} for artificial series generated by
the Random Midpoint Displacement (RMD) algorithm. The series have
length $N=2^{23}$ and Hurst exponent varying from $0.1$ to $0.9$
with step $0.1$. The parameter $\theta$ is taken equal to $0$,
$0.5$ and $1$ respectively.}
\end{figure}

\begin{figure}
\includegraphics[width=9.5cm,height=8cm,angle=0]{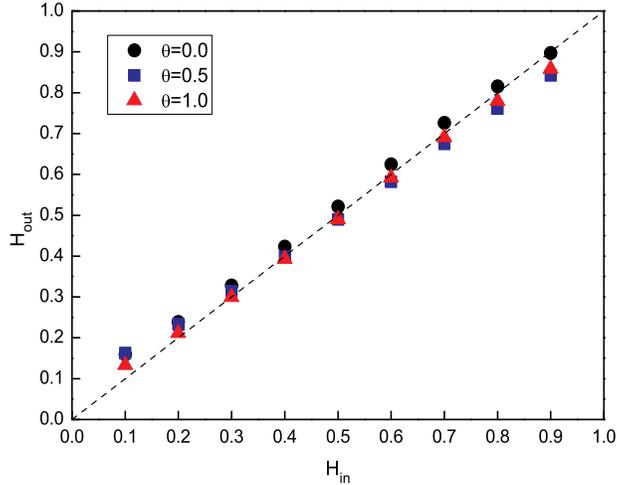}
\caption{\label{dma1} Values of the slopes of the function
$\sigma_{DMA}$ plotted in Fig.(\ref{dma0}) for three different
values of the parameter $\theta$ respectively equal to $0$, $0.5$
and $1$.}
\end{figure}

\begin{figure}
\includegraphics[width=9.5cm,height=8cm,angle=0]{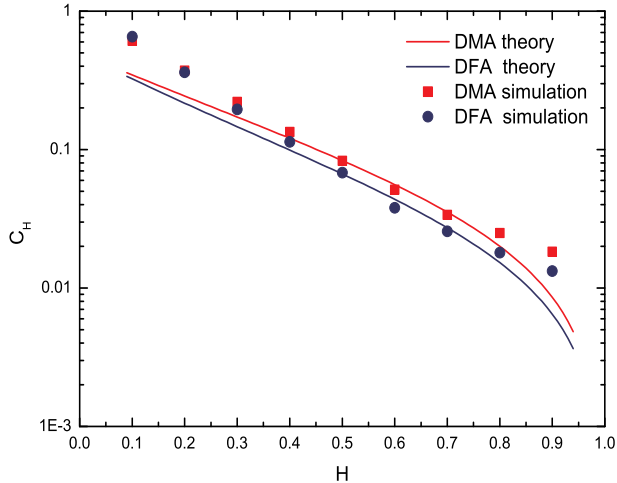}
\caption{\label{dma2} Values of $C_H$ for the DMA algorithm for
$\theta=0.5$ (squares) and for the DFA-1 algorithm applied to the
same series (circles). The solid lines represent the values of
$C_H$ calculated by using the expressions (\ref{CH}) and
(\ref{CHdfa1}) respectively.  }
\end{figure}

 In Fig.(\ref{dma0}), the results of the DMA algorithm
implemented over artificial fractional random walks generated by
the random midpoint displacement algorithm are shown. The length
of the walks  is $N=2^{23}$ and the Hurst exponent ranges from
$0.1$ to $0.9$ with step $0.1$. The curves in the three figures
refer respectively to three values of the parameter $\theta$,
namely $\theta=0$, $\theta=0.5$ and $\theta=1$.  The slopes of the
curves plotted in Fig.(\ref{dma0}) are plotted in
Fig.(\ref{dma1}). The slopes and the intercepts of the curves
plotted in Fig.(\ref{dma0}) are reported in Table I.  From the
data shown in Table I, it is possible to deduce that the DMA with
$\theta=0.5$ performs better with positively correlated signals
with $0.5<H<1$, while the DMA with $\theta=0$ and $\theta=1$
performs better with negatively correlated signals with $0<H<0.5$.
In Fig.(\ref{dma2}), the theoretical values of $C_H$, calculated
by using the Eq.(\ref{CH}), are compared with those obtained by
the intercepts of the curves plotted in Fig.(1) for $\theta=0.5$
(data of the 2nd column of Table I \cite{note}).

\begin{table*}
\caption{\label{tab:table1} Intercept (A) and slope (B) for the
curves plotted in Fig.1. }
\begin{ruledtabular}
\begin{tabular}{ccccccc}
&\multicolumn{2}{c}{$\theta=0$}&\multicolumn{2}{c}{$\theta=0.5$}&\multicolumn{2}{c}{$\theta=1$}\\
  $H$ & A & B & A & B & A & B\\ \hline

 $0.1$ & -0.73211  & 0.13220 & -0.84280  & 0.14151 & -0.71759  & 0.12861\\
 $0.2$ & -1.48755  & 0.21675 & -1.62576  & 0.22023 & -1.47214  & 0.21294 \\
 $0.3$ & -2.23912   & 0.30962 & -2.41792   & 0.30952 & -2.22270   & 0.30555\\
$0.4$ &  -2.99192 & 0.40900 &  -3.20887 & 0.40212 &  -2.97326 & 0.40421\\
 $0.5$ & -3.72180  & 0.50752 & -4.00245 & 0.50058 & -3.70336 & 0.50301\\
 $0.6$ & -4.45987  & 0.61307  & -4.80327  & 0.60187 & -4.44063  & 0.60843 \\
$0.7$ & -5.17084  & 0.71296 & -5.60555  & 0.70333 & -5.14974  &0.70776\\
$0.8$ & -5.84043  & 0.79625  & -6.40763  & 0.79829 & -5.82297  & 0.79277\\
 $0.9$ & -6.51500  & 0.87257 & -7.27298  & 0.89698 & -6.49215  & 0.86705 \\

\end{tabular}

\end{ruledtabular}
\end{table*}

It is interesting a comparison of the Eqs. (\ref{CH0},\ref{CH05})
with the corresponding ones  obtained for the DFA method, that we
will briefly recall here. According to the DFA method, the
integrated profile $y(i)$ is divided into boxes of equal
 length $n$. In each box, the signal $y(i)$ is best-fitted by an $\ell$-order  polynomial
 $y_{n,\ell}(i)$, which represents the local trend in that
 box.
 The different order of the DFA-$\ell$ (e.g.,DFA-0 if $\ell=0$, DFA-1 if $\ell=1$, DFA-2 if $\ell=2$, etc)
 is obtained according to the order of the polynomial fit.
Finally, for each box $n$, the variance:
\begin{equation}
\label{DFA_Fn}
  \sigma_{DFA}^2\equiv {\frac{1}{N}\sum_{i=1}^{N}[y(i)-y_{n}(i)]^{2}}
\end{equation}
\noindent is calculated.
 The calculation is then repeated
for different box lengths $n$, yielding  the behavior of
$\sigma_{DFA}$ over a broad range of scales. For scale-invariant
signals with power-law correlations, the following relationship
between the function $\sigma_{DFA}$ and the scale $n$ holds:
\begin{equation}
\label{DFA}
 \sigma_{DFA}^2 \sim n^{2H}.
\end{equation}

 The asymptotic behavior of the DFA-0 and DFA-1
functions has been investigated by Raymond and Bassingthwaighte in
\cite{Raymond} and by Taqqu, Teverovsky and Willinger in
\cite{Taqqu} respectively. Raymond and Bassingthwaighte worked out
the following relation (Eq.(21) of Ref. \cite{Raymond}) for the
asymptotic scaling of the $\sigma_{DFA-0}$ function \cite{note2}:

\begin{equation}
\label{CHdfa0}
  \sigma_{DFA-0}^2
\simeq\left(\frac{1}{2H+1}-\frac{1}{2(H+1)}\right) n^{2H}
\end{equation}

The asymptotic behavior of the DFA-1 function is reported in the
Appendix of \cite{Taqqu}:
\begin{equation}
\label{CHdfa1}
 \sigma_{DFA-1}^2
\simeq\left(\frac{2}{2H+1}+\frac{1}{H+2}-\frac{2}{H+1}\right)
n^{2H}
\end{equation}

In Fig.~\ref{dma2}, the values of $C_H$ for the DMA and the DFA
are shown. It can be observed that the behavior of $C_H$ obtained
from the simulations (square and circles) follows quite well the
analytical curves (solid lines) around $H\simeq 0.5$. Deviations
are observed at the extrema of the $H$ range. Such deviations
might be related either to the DFA and DMA sensitivity or to the
RMD accuracy. This issue, which is beyond the scope of the present
effort, will be addressed in a future paper.

\section{Conclusions}

We have analytically derived the asymptotic scaling behavior  of
the DMA algorithm proposed in \cite{Alessio}. We have compared the
 analytical results with those obtained from the simulations of
fractional Brownian paths with assigned values of $H$ obtained by
the Random Midpoint Displacement.


\begin{thebibliography}{99}
\bibitem{note} Since  the Fractional Brownian motions, used for the simulations
plotted in Fig. (1), have been generated by the Random Midpoint
Displacement algorithm. In the calculation of $C_H$ it must be
kept in mind that: $\sigma^2_{RMD}= {{(1-2^{2H-2})}}/{2^{2H \nu}}
\sigma^2_{gauss} \hspace*{25 pt} $ where $\nu$ is the number of
steps of the RMD algorithm.
\end{thebibliography}
\end{document}